\documentclass[%
 reprint,
 amsmath,amssymb,
 aps,
 floatfix,
]{revtex4-2}

\usepackage{graphicx}
\usepackage{dcolumn}
\usepackage{bm}
\usepackage{layout}
\usepackage{stfloats} 
\usepackage{float}
\usepackage{placeins}
\usepackage{booktabs}
\usepackage{caption} 

\begin{document}

\preprint{APS/123-QED}

\title{Description of Charged\text{-}Particle Multiplicity Distributions in High\text{-}Energy Proton\text{-}Proton Collisions Based on a Two-Component Model and Examination of Parton Distribution Functions}

\author{Zhixiang Yang}
\affiliation{%
  Department of Physics, East China Normal University\\
  Shanghai 200241, People's Republic of China
}

\author{Jianhong Ruan}
\email{Corresponding author: jhruan@phy.ecnu.edu.cn}
\affiliation{%
  Department of Physics, East China Normal University\\
  Shanghai 200241, People's Republic of China
}

\date{\today}

\begin{abstract}
High-energy proton-proton collisions at the LHC offer a stringent test of Quantum Chromodynamics (QCD) in the small-$x$, gluon-dominated regime. 
This study focus on a minimal, gluon-driven framework to describe the charged-particle multiplicities and their pseudorapidity 
densities in high energy collisions. The two-component model taken here includes the hard
gluon-gluon fusion process and the soft quark recombination process, which directly  relates to both integrated and unintegrated parton distributions. We begin by evolving Parton Distribution Functions (PDFs) using the Modified Dokshitzer-Gribov-
Lipatov-Altarelli-Parisi (MD-DGLAP) equations. These PDFs are then converted 
into unintegrated PDFs (UPDFs) via the Kimber-Martin-Ryskin (KMR) skeme. The resulting PDFs and UPDFs are incorporated into the two-component model 
to predict the charged-particle pseudorapidity density $\left(1 / N_{\mathrm{ev}}\right) d N_{\mathrm{ch}} / d \eta$ in $pp$ collisions at LHC 
energies. Our predictions are compared to the data from the ATLAS experiment, revealing that the model effectively captures the features
of the observed pseudorapidity distributions, despite its simplicity. Within this framework, the gluon-gluon fusion processes are found to dominate particle production for  $\sqrt { s } \ge 9 0 0 \ \mathrm { GeV }$.These findings provide phenomenological support for MD-DGLAP-based PDFs and the associated small-$x$ gluon dynamics. Furthermore,a comparative analysis of results from alternative PDF sets—including CTEQ, MSHT, NNPDF, HERAPDF, and GRV—is performed, with particular focus on examining their consistency with the relative shapes of experiment data in the small-$x$ region.
\end{abstract}

\maketitle

\section{INTRODUCTION}

High-energy proton-proton ($pp$) collisions provide a crucial experimental arena for testing parton distribution functions (PDFs). Owing to the 
highly complex and nonperturbative nature of hadronic interactions, however, experimental observables do not straightforwardly reflect the 
underlying partonic structure of the proton. As a result, a variety of theoretical models have been developed to describe particle production 
in high-energy $pp$ collisions, including the string fragmentation model\cite{andersson1983parton}, 
the Parton Cascade plus Hydrodynamic Model \cite{werner2010event}, 
and frameworks based on the Color Glass Condensate (CGC) coupled with hydrodynamics~\cite{gelis2010color}. Most of these approaches rely heavily on phenomenological 
parameterizations and do not explicitly clarify how the pre-collision parton distributions inside the proton are mapped onto the post-collision 
energy deposition and particle production processes.

Some studies~\cite{kharzeev2005color,albacete2007particle} have described particle production using a single mechanism, namely gluon--gluon 
fusion, $gg\to g$, as proposed by Gribov, Levin, and Ryskin (GLR)~\cite{gribov1983semihard}. However, the single-inclusive spectrum exhibits a 
rapidity distribution with three characteristic domains: a central region and two fragmentation regions associated with the projectile and the 
target~\cite{ochs1977hadron}. The GLR mechanism is expected to be most relevant near midrapidity, implying that additional dynamics are required 
to account for particle production in the fragmentation regions.
This observation motivates a two-component description of hadronic collisions, such as the HIJING model~\cite{wang1991hijing,wang1997pqcd}. 
HIJING employs parametrized collinear (integrated) parton distributions to compute multiple minijet production and uses the Lund string 
model~\cite{andersson1987model,nilsson1987interactions} to describe soft beam-jet fragmentation. Nevertheless, minijet production in the central 
region is more naturally formulated in terms of unintegrated gluon distributions rather than collinear gluon 
densities~\cite{szczurek2003unintegrated}, while the string-fragmentation component is not directly linked to the underlying parton distributions.
In Ref.~\cite{ruan2010particle}, a two-component framework distinct from HIJING was proposed, combining a hard gluon--gluon fusion 
mechanism~\cite{gribov1983semihard} with a soft quark recombination process~\cite{das1977quark,hwa1980clustering}. In this picture, at sufficiently 
high collision energies, particles originating from hard gluon fusion populate the midrapidity region, with the initial gluons described by the 
unintegrated gluon distributions of the colliding hadrons. Meanwhile, valence quarks largely retain their original collinear distributions and 
hadronize via recombination with additional samll-$x$ sea quarks from the central region. Consequently, the model requires both integrated and 
unintegrated parton distributions.

In this work, we adopt the two-component model proposed in Ref.~\cite{ruan2010particle}. At the time of that study, experimental data at 
center-of-mass energies above $\sqrt{s}\geq 900\ \mathrm{GeV}$ were not yet available, and the analysis primarily focused on providing 
theoretical predictions. By contrast, the present work applies this model to existing experimental measurements 
at $\sqrt{s}\geq 900\ \mathrm{GeV}$, performs a systematic comparison of results obtained with different parton distribution functions, and 
thereby aims to improve the understanding of the phenomenological characteristics of the currently available PDF sets.

Parton distribution functions (PDFs) constitute essential ingredients of the quantum chromodynamics (QCD) factorization framework and encode 
the momentum distributions of quarks and gluons inside the proton~\cite{collins1989perturbative,collins2011foundations}. They are commonly 
classified into two categories: integrated (collinear) parton distributions (PDFs) and unintegrated parton distribution functions (UPDFs). Generally, the term “PDF” 
refers to the integrated distribution, which can be obtained by integrating the corresponding UPDF over the parton transverse momentum. 
Integrated PDFs are primarily employed in the description of inclusive processes, whereas UPDFs retain an explicit dependence on the 
transverse momentum and are therefore suited for the treatment of exclusive and semi-exclusive 
reactions~\cite{collins2011foundations,kimber2001unintegrated}.

Several research collaborations, including CTEQ~\cite{yan2023ct18}, MSHT~\cite{bailey2021parton}, NNPDF~\cite{ball2015parton}, 
HERAPDF~\cite{collaborations2015combination}, and GRV~\cite{gluck1998dynamical} regularly update their PDF sets through global QCD analyses. 
The most significant differences among these sets are observed in the gluon distribution, particularly in the small-$x$ region, where gluons 
dominate the proton structure while experimental constraints remain relatively weak~\cite{gao2018structure}. These discrepancies highlight 
the theoretical challenges associated with the small-$x$ region and motivate the development of evolution equations that extend beyond the 
standard Dokshitzer-Gribov-Lipatov-Altarelli-Parisi (DGLAP) framework~\cite{altarelli1977asymptotic}.

In contrast to integrated PDFs, unintegrated parton distribution functions (UPDFs) are considerably less well constrained. The main difficulty 
arises from their nontrivial dependence on at least two variables, $x$ and $k^2$, for which no universal analytical form exists. An additional 
challenge concerns the separation of perturbative and nonperturbative contributions. Moreover, different physical processes focus on UPDFs in 
distinct regions of phase space. Consequently, various UPDF models proposed in the literature—such as those based on the 
BFKL~\cite{kuraev1977pomeranchuk}, GBW~\cite{golec1998saturation}, and KL~\cite{kharzeev2001manifestations} approaches—exhibit substantial 
differences in both formulation and phenomenological behavior.

The Kimber-Martin-Ryskin (KMR) scheme~\cite{kimber2001unintegrated} provides a well-established procedure for constructing unintegrated 
parton distribution functions (UPDFs) from conventional collinear PDFs. In this approach, a kinematic constraint is imposed on the final step 
of the parton evolution, and the Sudakov form factor is introduced to account for the probability of no parton emission between two scales. 
As a result, an explicit dependence on a second scale, $\mu^2$, is generated. The resulting unintegrated distributions effectively resum 
large logarithmic contributions of $\ln(1/x)$ within a DGLAP-based evolution framework, rendering the KMR formalism particularly 
suitable for the description of small-$x$ kinematics.

In the present work, we construct UPDFs using the KMR scheme, where a key issue concerns the choice of the input collinear PDFs. Conventional 
PDFs evolve according to the linear DGLAP equations, which predict a rapid growth of the gluon distribution in the small-$x$ region, where 
gluon-gluon recombination can no longer be neglected. Corrections due to gluon recombination were first proposed by Gribov, Levin, and 
Ryskin~\cite{gribov1983semihard}, and later formulated more explicitly by Mueller and Qiu in the GLR-MQ equation~\cite{mueller1986gluon}. 
Subsequently, a modified DGLAP (MD-DGLAP) evolution equation was developed~\cite{zhu1999new} to remedy a deficiency of the GLR-MQ 
approach, namely its violation of momentum conservation arising from the neglect of antishadowing effects. In Ref.~\cite{ruan2009prediction}, 
the MD-DGLAP formalism was extended down to an initial scale of $Q_0^2 = 0.34\ \mathrm{GeV}^2$ and shown to provide a good description of 
structure function data. In this work, we adopt the parton distributions obtained in that study as inputs to the KMR scheme. Furthermore, 
within the KMR construction employed here, the corresponding gluon recombination corrections are consistently taken into 
account~\cite{ruan2009prediction}.

The remainder of this paper is organized as follows. In Section II, we present the theoretical framework for parton distributions: we first 
briefly review the MD-DGLAP evolution equations and then describe the KMR prescription for constructing unintegrated distributions from the 
resulting collinear PDFs. In Section III, these parton distributions are applied to the analysis of charged-particle production in 
high-energy $pp$ collisions. We introduce the two-component model to calculate multiplicity distribution spectra and compare the results 
with LHC data, including a systematic assessment of scenarios with and without valence-quark contributions. The distinct roles of gluon- 
and quark-initiated processes in different kinematic regions are discussed. We also compare our calculations based on CTEQ, MSHT, NNPDF, 
HERAPDF, and GRV parton distributions and analyze their differing phenomenological characteristics in relation to experimental data. 
Finally, Section IV summarizes the main results and discusses the examination of PDFs in the future experimental studies.

\section{Parton Distribution Functions}

\subsection{MD-DGLAP evolution}

The standard Dokshitzer-Gribov-Lipatov-Altarelli-Parisi (DGLAP) evolution equations~\cite{altarelli1977asymptotic} provide a successful 
description of the $Q^2$ dependence of parton distribution functions over a wide kinematic range. In the small-$x$ region, however, the 
linear DGLAP evolution predicts a rapid, power-like growth of the gluon density, which eventually leads to a violation of unitarity due 
to the neglect of gluon recombination effects~\cite{gribov1983semihard,mueller1986gluon}. To restore unitarity and incorporate higher-twist 
contributions associated with gluon recombination processes ($gg \to g$), several nonlinear extensions of the DGLAP framework have been 
proposed~\cite{gribov1983semihard,mueller1986gluon,zhu1999new}.

In the present work, we adopt the modified DGLAP (MD-DGLAP) evolution equations developed in Refs.~\cite{zhu1999new,ruan2009prediction}, 
which consistently incorporate both shadowing and antishadowing effects and thereby preserve momentum conservation. Within this framework, 
the evolution equations for the gluon distribution $xg(x,Q^2)$ and the sea-quark distribution $xs(x,Q^2)$ are modified to include nonlinear 
recombination corrections, while the valence-quark distributions continue to evolve according to the conventional linear DGLAP equations.

The evolution equation for the gluon distribution can be written as:

\begin{widetext}
\begin{equation}
\begin{split}
Q^{2}\frac{d\,x g(x,Q^{2})}{dQ^{2}}
&= \frac{\alpha_s(Q^{2})}{2\pi} \int_x^1 \frac{dy}{y}
   \Bigl[ P_{gq}(z)\, x s(y,Q^{2})
        + P_{gg}(z)\, x g(y,Q^{2}) \Bigr] \\
&\quad - \frac{\alpha_s(Q^{2})}{2\pi} x g(x,Q^{2})
   \Biggl[ n_f \int_0^1 dz\, P_{qg}(z)
         + \frac{1}{2} \int_0^1 dz\, P_{gg}(z) \Biggr] \\
&\quad - \frac{\alpha_s^2(Q^{2}) K}{Q^{2}} \int_x^{1/2} \frac{dy}{y}\,
   x P_{gg\to g}(z) \bigl[y g(y,Q^{2})\bigr]^2 \\
&\quad + \frac{\alpha_s^2(Q^{2}) K}{Q^{2}} \int_{x/2}^x \frac{dy}{y}\,
   x P_{gg\to g}(z) \bigl[y g(y,Q^{2})\bigr]^2 .
\end{split}
\label{eq:1}
\end{equation}
\end{widetext}

where $z = x/y$. The corresponding sea quark evolution equation is

\begin{widetext}
\begin{equation}
\begin{split}
Q^{2}\frac{d\,x s(x,Q^{2})}{dQ^{2}}
&= \frac{\alpha_s(Q^{2})}{2\pi} \int_x^1 \frac{dy}{y}
   \Bigl[ P_{qq}(z)\, x s(y,Q^{2})
        + P_{qg}(z)\, x g(y,Q^{2}) \Bigr] \\
&\quad - \frac{\alpha_s(Q^{2})}{2\pi} x s(x,Q^{2)}
   \cdot \frac{1}{2} \int_0^1 dz\, P_{qq}(z) \\
&\quad - \frac{\alpha_s^2(Q^{2}) K}{Q^{2}} \int_x^{1/2} \frac{dy}{y}\,
   x P_{gg\to q}(z) \bigl[y g(y,Q^{2})\bigr]^2 \\
&\quad + \frac{\alpha_s^2(Q^{2}) K}{Q^{2}} \int_{x/2}^x \frac{dy}{y}\,
   x P_{gg\to q}(z) \bigl[y g(y,Q^{2})\bigr]^2 .
\end{split}
\label{eq:2}
\end{equation}
\end{widetext}

In Eqs.~\eqref{eq:1} and \eqref{eq:2}, the first two lines correspond to the conventional leading-order DGLAP evolution, involving the 
standard Altarelli-Parisi splitting functions $P_{ij}(z)$. The remaining two terms represent nonlinear corrections arising from gluon 
recombination effects. These corrections enter with opposite signs: the negative contribution describes the depletion of parton densities 
due to shadowing through $gg \to g$ (or $gg \to q\bar{q}$ in the case of sea quarks), whereas the positive contribution accounts for the 
corresponding enhancement at larger values of $x$, commonly referred to as antishadowing.

The explicit form of the modified DGLAP (MD-DGLAP) equations, including the recombination functions $P_{gg\to g}(z)$ and $P_{gg\to q}(z)$ 
associated with gluon recombination processes, follows Refs.~\cite{zhu1999new,ruan2009prediction}. These works also provide the 
phenomenological motivation for the correlation parameter $K$, which characterizes the transverse overlap of the two recombining partons 
and determines the normalization of the double-parton distribution. In accordance with the value obtained from the global fit to HERA data 
in Ref.~\cite{ruan2009prediction}, we adopt $K=0.0014 \ \mathrm{GeV}^2$. For completeness, both the standard splitting functions and the 
recombination functions are listed in Appendix~\ref{app:splitting}.

The initial parton distributions at the input scale $Q_0^2 = 0.34 \ \mathrm{GeV}^2$ are taken from the parametrization of 
Ref.~\cite{ruan2009prediction}. They are assumed to follow a general valence-like form, $xg(x,Q_0^2) = A_g x^{B_g} (1-x)^{C_g}$ 
for the gluon distribution, with analogous expressions for the quark distributions, as commonly adopted in dynamical PDF 
approaches~\cite{gluck1998dynamical}. The specific parameter values for all parton flavors are summarized in 
Appendix~\ref{app:splitting}.

\subsection{Unintegrated parton distributions}
\label{sec:uPDF}

Within the Kimber-Martin-Ryskin (KMR) framework~\cite{kimber2001unintegrated}, unintegrated parton distribution functions (UPDFs) are 
constructed from conventional collinear PDFs by explicitly separating the dependence on the parton transverse momentum $k_t$ from that on 
the factorization scale $\mu$. This separation is implemented by retaining only the real emissions in the final step of the DGLAP evolution, 
while resumming virtual corrections into a Sudakov form factor. The Sudakov factor represents the probability of no resolvable parton 
emission between the scales $k_t$ and $\mu$, thereby ensuring the correct ordering of emissions.

Following the KMR prescription and extending it to incorporate the nonlinear corrections arising from the MD-DGLAP evolution, the gluon UPDF 
can be written as

\begin{widetext}
\begin{equation}
\begin{split}
f_g(x,k_t^2,\mu^2) &= T_g(k_t^2,\mu^2)
\Bigg\{
\frac{\alpha_s(k_t^2)}{2\pi} \int_x^{z_{\max}} dz \Bigg[
P_{gg}(z) \frac{x}{z} g\left(\frac{x}{z},k_t^2\right)
+ P_{gq}(z) \frac{x}{z} \Bigl(V+S\Bigr)\left(\frac{x}{z},k_t^2\right)
\Bigg] \\
&\quad - \frac{\alpha_s^2(k_t^2) K}{k_t^2} \int_x^{1/2} \frac{dy}{y}\, x P_{gg\to g}(z)
\left[y g(y,k_t^2)\right]^2 \\
&\quad + \frac{\alpha_s^2(k_t^2) K}{k_t^2} \int_{x/2}^x \frac{dy}{y}\, x P_{gg\to g}(z)
\left[y g(y,k_t^2)\right]^2
\Bigg\},
\label{eq:kmr_gluon}
\end{split}
\end{equation}
\end{widetext}
where $V(x,k_t^2)$ and $S(x,k_t^2)$ denote the valence and sea quark distributions, respectively. 
The Sudakov form factor $T_g(k_t^2,\mu^2)$, which resums virtual corrections and ensures no real emission in the region 
$k_t < k_t' < \mu$, reads
\begin{multline}
T_g(k_t^2,\mu^2)
=
\exp\Bigg\{
- \int_{k_t^2}^{\mu^2}
\frac{dk_t^{\prime 2}}{k_t^{\prime 2}}\,
\frac{\alpha_s(k_t^{\prime 2})}{2\pi}
\\
\times
\Bigg[
\frac{1}{2}
\int_{z_{\min}}^{z_{\max}} dz\, P_{gg}(z)
+ n_f \int_0^1 dz\, P_{qg}(z)
\Bigg]
\Bigg\} .
\end{multline}

The upper limit of the $z$ integration, $z_{\max} = \mu/(\mu + k_t)$, follows from the strong $k_t$-ordering condition imposed on the final 
emission step and effectively implements angular ordering, thereby suppressing soft-gluon coherence effects~\cite{kimber2001unintegrated}. 
The corresponding lower limit is given by $z_{\min} = 1 - z_{\max}$.

The nonlinear contributions, represented by the second and third lines in Eq.~\eqref{eq:kmr_gluon}, are incorporated in accordance with 
the structure of the MD-DGLAP evolution, thus preserving gluon recombination effects at the level of unintegrated distributions. In these terms, the scale appearing in the denominator is taken to be $k_t^2$, rather than $Q^2$, reflecting the transverse momentum of the parton emitted in the final evolution step.

\begin{figure}[H]
    \centering
    \includegraphics[width=\columnwidth]{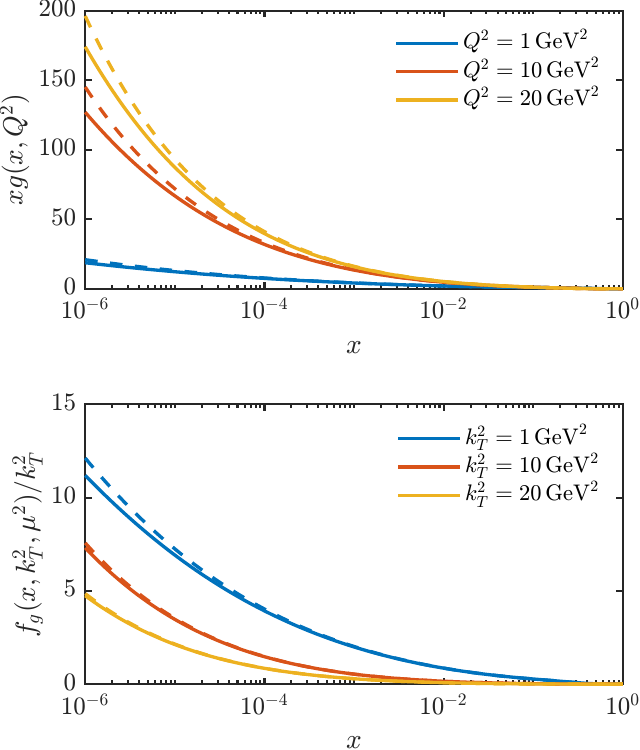}
    \caption{Integrated and unintegrated gluon distributions.\textbf{Top}: the integrated gluon distribution functions $xg(x,Q^{2})$, 
    evolved using the MD-DGLAP equations (solid lines) and DGLAP equations (dashed lines)(shown for \(Q^{2}=1,\,10,\,20~\mathrm{GeV}^{2}\)). 
    \textbf{Bottom}: the corresponding gluon unintegrated distribution functions , \(f_{g}(x,k_{T}^{2},\mu^{2})/k_{T}^{2}\), obtained from 
    the PDFs via the KMR prescription with higher-twist corrections (solid lines) and without higher-twist corrections (dashed lines) at 
    \(\mu^{2}=100~\mathrm{GeV}^{2}\) (shown for \(k_{T}^{2}=1,\,10,\,20~\mathrm{GeV}^{2}\)). }
    \label{fig:gluon_dist}
\end{figure}

For illustration, Fig.~\ref{fig:gluon_dist} displays the collinear gluon distribution $xg(x,Q^2)$ obtained from the MD-DGLAP evolution 
(upper panel) together with the corresponding unintegrated gluon distribution $f_g(x,k_t^2,\mu^2)$ constructed via the KMR procedure 
at $\mu^2 = 100\ \mathrm{GeV}^2$ (lower panel)(solid lines). The unintegrated distribution is presented in the form $f_g(x,k_t^2,\mu^2)/k_t^2$ to 
facilitate comparison with the collinear limit, in which integration over $k_t^2$ up to $\mu^2$ reproduces the integrated PDF. 
To further 
highlight the impact of higher-twist effects, Fig.~\ref{fig:gluon_dist} also shows the corresponding integrated and unintegrated gluon 
distributions obtained without higher-twist corrections(dashed lines). As can be seen, in the small-$x$ region, the higher-twist effect effectively suppresses the rapid growth of the gluon distribution.

\section{Interpretation of Particle Multiplicity Production in High-Energy Collisions}

\subsection{A two-component model}

High-energy $pp$ collisions exhibit a characteristic rapidity structure in inclusive hadron production, consisting of a central region and two fragmentation regions associated with the projectile and the target. In the central region, characterized by small values of $x$ and semi-hard transverse momenta, particle production is dominated by gluon-initiated processes, leading predominantly to the formation of light mesons such as pion and kaon. In contrast, hadron production in the fragmentation regions is mainly governed by valence-quark dynamics and can be effectively described within the quark recombination framework~[14].

A key advantage of the present approach is that it enables a unified analysis of experimental data across different rapidity regions through the consistent use of both integrated and unintegrated parton distribution functions. This feature allows the underlying partonic mechanisms responsible for particle production in distinct kinematic domains to be treated within a common framework, which is difficult to achieve in many existing phenomenological models.

\textbf{Component I(central production).} At sufficiently high energies, the cross section for inclusive gluon production in $pp \to g$ via the gluonic 
subprocess $gg \to g$ is given by:
	\begin{equation}
	\begin{array} { l } { { \displaystyle \frac { d \sigma _ { p - p } ^ { I } ( y_g , \, p _ { t , g } ) } { d y_g \, d ^ { 2 } p _ { t , g } } = \frac { 4 N _ { c } } { N _ { c } ^ { 2 } - 1 } \frac { 1 } { p _ { t , g } ^ { 2 } } \int d ^ { 2 } q _ { t , g } \alpha _ { s } ( \Omega ) } } \\ { { \displaystyle \qquad \qquad \qquad \times \, \mathcal{F} _ { g } ^ { p } \left( x _ { 1 } , \, \left( \frac { p _ { t , g } + q _ { t , g } } { 2 } \right) ^ { 2 } , \, p _ { t , g } ^ { 2 } \right) } } \\ { { \displaystyle \qquad \qquad \times \, \mathcal{F} _ { g } ^ { p } \left( x _ { 2 } , \, \left( \frac { p _ { t , g } - q _ { t , g } } { 2 } \right) ^ { 2 } , \, p _ { t , g } ^ { 2 } \right) , } } \end{array}	
   \label{eq:gluonfusion}
	\end{equation}
where $x_{1,2} = \frac{p_{t,g}}{\sqrt{s}} \exp(\pm y_g)$, $k_{1t}^2 = \frac{1}{4}(p_{t,g} + q_{t,g})^2$, $k_{2t}^2 = \frac{1}{4}(p_{t,g} - q_{t,g})^2$, 
and $\Omega = \max(k_{1t}^2, k_{2t}^2, p_{t,g}^2)$. The two-scale unintegrated gluon distribution $\mathcal{F}_g(x,k_t^2,p_t^2)$ is 
related to the UPDF $f_g(x,k_t^2,\mu^2)$ introduced in Sec.~\ref{sec:uPDF} by $\mathcal{F}_g = f_g/k_t^2$.

Following the approach in Ref.~\cite{szczurek2003unintegrated}, the single-hadron inclusive spectrum is obtained by convoluting the 
gluon spectrum with a fragmentation function $D_{g\to h}(z,\mu_D^2)$:

\begin{equation}
\begin{split}
\frac{d\sigma(\eta_h, p_{t,h})}{d\eta_h \, d^2 p_{t,h}}
&= \int_{z_{\min}}^{z_{\max}} dz\,
\frac{J^2 D_{g\to h}(z, \mu_D^2)}{z^2} \\
&\quad \times
\left.
\frac{d\sigma(y_g, p_{t,g})}{dy_g \, d^2 p_{t,g}}
\right|_{y_{t,g} = \eta_h / z} \;
\end{split}
\end{equation}
where the Jacobian 
\begin{equation}
J(m_{t,h}, \eta_h) = \left(1 - \frac{m_h^2}{m_{t,h}^2 \cosh^2 \eta_h}\right)^{-1/2}   
\end{equation}
accounts for the 
transformation from rapidity to pseudorapidity, and $m_{t,h} = \sqrt{m_h^2 + p_{t,h}^2}$ is the transverse mass. The fragmentation 
function is taken from Ref.~\cite{gao2024global}, valid for $\mu_D\ge 4$GeV. For lower scales ($p_{t,h}<4$~GeV), we extrapolate using 
the scaling $D_{g\to h}(z,\mu_D^2) = \sqrt{\mu_D/4} \ D_{g\to h}(z,16\,\mathrm{GeV}^2)$, consistent with the behavior observed in 
phenomenological studies.

The charged-particle pseudorapidity density in the central region (Component I) is then
\begin{equation}
\begin{aligned}
\frac{dN_{pp}^{I}}{d\eta} 
&= C(\sqrt{s})
   \frac{d\sigma(\eta_h)}{d\eta_h}  \\
&= C(\sqrt{s})
   \int d^{2}p_{t,h}\;
   \frac{d\sigma(\eta_h, p_{t,h})}
        {d\eta_h\, d^{2}p_{t,h}} \; ,
\end{aligned}
\label{eq:mult_I}
\end{equation}
where the coefficient $C(\sqrt{s})$ is hard to be obtained in parton level theoretically since it contains the complex nonperturbative informations.  
The factor $C(\sqrt{s})$ is adjusted to match the data at $\eta=0$ (see Sec.~\ref{sec:results} for details).
The integral here includes a sum over pions and kaons
as the primary contributions from gluon-initiated processes. 

\textbf{Component II (fragmentation regions).} According to the quark recombination model~\cite{das1977quark,hwa1980clustering}, 
the valence quarks of incident protons tend to fly through the central region with their original momentum fraction. These valence 
quarks recombine with low-$p_t$ antiquarks to produce outgoing hadrons in the fragmentation region. The cross section for proton 
production in $pp$ collisions in this model is
	\begin{equation}
	\begin{array} { l } { { \displaystyle \frac { 1 } { \sigma _ { \mathrm { i n } } } \frac { d \sigma _ { p - p } ^ { I I } } { d x \, d p _ { t } ^ { 2 } } = 6 \frac { 1 - x } { x } \int _ { 0 } ^ { x } d x _ { 1 } x _ { 1 } v _ { p } \big ( x _ { 1 } , p _ { t } ^ { 2 } \big ) } } \\ { { \displaystyle \qquad \qquad \qquad \times \frac { 1 } { 2 } ( 1 + \delta ) ( x - x _ { 1 } ) s _ { p } \big ( x - x _ { 1 } , p _ { t } ^ { 2 } \big ) , } } \end{array}
	\end{equation}
where $\delta s _ { p } ( x , p _ { t } ^ { 2 } )$ is the distribution of additional sea quarks in the central region and we assume that it has the form like $s _ { p } ( x , p _ { t } ^ { 2 } )$.  

The rapidity for protons in the recombination processes is introduced as
	\begin{equation}
	y = \ln x - \ln { \frac { \sqrt { m _ {p} ^ { 2 } + p _ { t } ^ { 2 } } } { \sqrt { s } } } .
	\end{equation}
where $m_p$ is the proton mass.Thus, we have
	\begin{equation}
	\frac { 1 } { \sigma _ { \mathrm { i n } } } \frac { d \sigma _ { p - p } ^ { I I } ( y , \, p _ { t } ) } { d y \, d p _ { t } ^ { 2 } } = \left. J _ { I I } ( y ; p _ { t } ; m _ { \pi } ) \frac { 1 } { \sigma _ { \mathrm { i n } } } \frac { d \sigma _ { p - p } ^ { I I } ( x , \, p _ { t } ) } { d x \, d p _ { t } ^ { 2 } } \right| _ { x \to y } ,
	\end{equation}
with a new Jacobian
	\begin{equation}
	J _ { I I } ( y ; p _ { t } ; m _ { \pi } ) = \frac { \partial x } { \partial y } = \frac { \sqrt { p _ { t } ^ { 2 } + m _ { \pi } ^ { 2 } } } { \sqrt { s } } e ^ { y } .
	\end{equation}
This leads to the pseudorapidity density
	\begin{equation}
	\begin{array} { l } { { \displaystyle \frac { d N _ { p - p } ^ { I I } } { d \eta } = \frac { 1 } { \sigma _ { \mathrm { i n } } } \int d p _ { t } ^ { 2 } \frac { d \sigma _ { p - p } ^ { I I } ( \eta , \, p _ { t } ) } { d \eta \, d p _ { t } ^ { 2 } } \ ~ } } \\ { { \displaystyle ~ ~ ~ ~ ~ ~ ~ ~ = \frac { 1 } { \sigma _ { \mathrm { i n } } } \int d p _ { t } ^ { 2 } J _ { I I } ( \eta ; p _ { t } ; m _ { \pi } ) \left. \frac { d \sigma _ { p - p } ^ { I I } ( y , \, p _ { t } ) } { d y \, d p _ { t } ^ { 2 } } \right\vert _ { y \to \eta } , } } \end{array}	
	\end{equation}
where we integrate over transverse momenta at $p_t<1\mathrm{GeV}$, since the recombination model works at lower $p_t$ 
range.

Summing the contributions of two components yields the total distribution
\begin{equation}
\frac{dN_{pp}}{d\eta} = \frac{dN_{pp}^{I}}{d\eta} + \frac{dN_{pp}^{II}}{d\eta} .
\label{eq:total_mult}
\end{equation}

The production mechanisms (I) and (II) utilize UPDFs and PDFs, respectively. Specifically, the PDFs are evolved from the MD-DGLAP 
equations as described in Sec.~2.1, while the UPDFs are calculated using the KMR mechanism presented in Sec.~2.2. We employ these 
parton distributions to predict the charged-particle multiplicity distributions.

\subsection{Description of $(1/N_{\mathrm{ev}}) \cdot (\mathrm{d}N_{\mathrm{ch}}/\mathrm{d}\eta)$ in proton-proton collisions}
\label{sec:results}

In this section, we compare the two-component model calculations of charged-particle production in proton-proton collisions with experimental data from the ATLAS Collaboration at center-of-mass energies of 0.9, 2.36, 7, and 13~TeV~\cite{atlas2010charged,atlas2016charged}. In particular, the prediction for $(1/N_{\mathrm{ev}}) \cdot (\mathrm{d}N_{\mathrm{ch}}/\mathrm{d}\eta)$ is evaluated against the corresponding ATLAS measurements under the kinematic selections $n_{\mathrm{ch}} \geq 1$ and $p_t > 500~\mathrm{MeV}$.

\begin{figure}[H]
    \centering
    \includegraphics[width=\columnwidth]{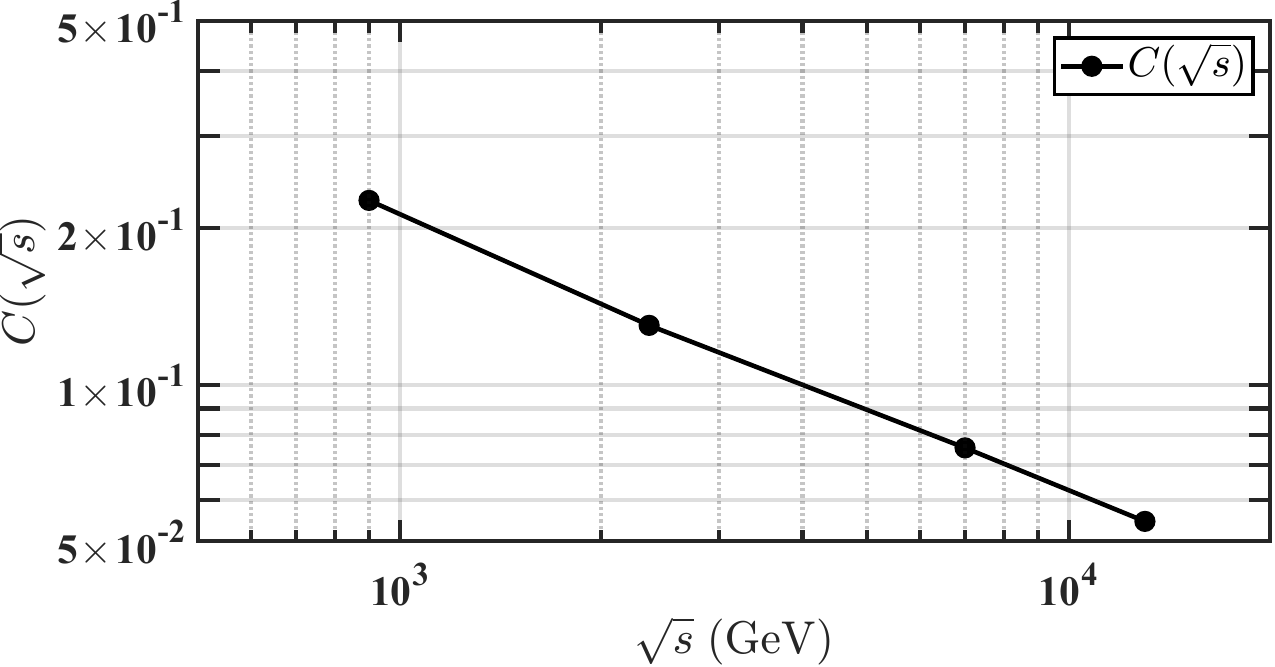}
    \caption{Normalization coefficient $C(\sqrt{s})$ in the central
    component of the two-component model. The solid line represents
    a linear fit in logarithmic scale.}
    \label{fig:coefficient}
\end{figure}

For the pseudorapidity distribution $(1/N_{\mathrm{ev}}) \cdot (\mathrm{d}N_{\mathrm{ch}}/\mathrm{d}\eta)$, the model reproduces the overall features of the experimental data at all considered energies, including the approximately flat behavior in the central region and the characteristic fall-off at large $|\eta|$. It should be noted that the coefficient $C(\sqrt{s})$ in the model is fixed using the experimental values at $\eta = 0$. Proton--proton collisions involve highly complex nonperturbative dynamics, and the extracted values of $C(\sqrt{s})$ obtained from the data are shown in Fig.~\ref{fig:coefficient}. These values are found to approximately satisfy the following empirical relation:

\begin{equation}
\mathrm{ln}(C(\sqrt{s}))=-0.528\mathrm{ln}(\sqrt{s})+2.089
\end{equation}

\begin{figure*}[t]
    \centering
    \includegraphics[width=\textwidth]{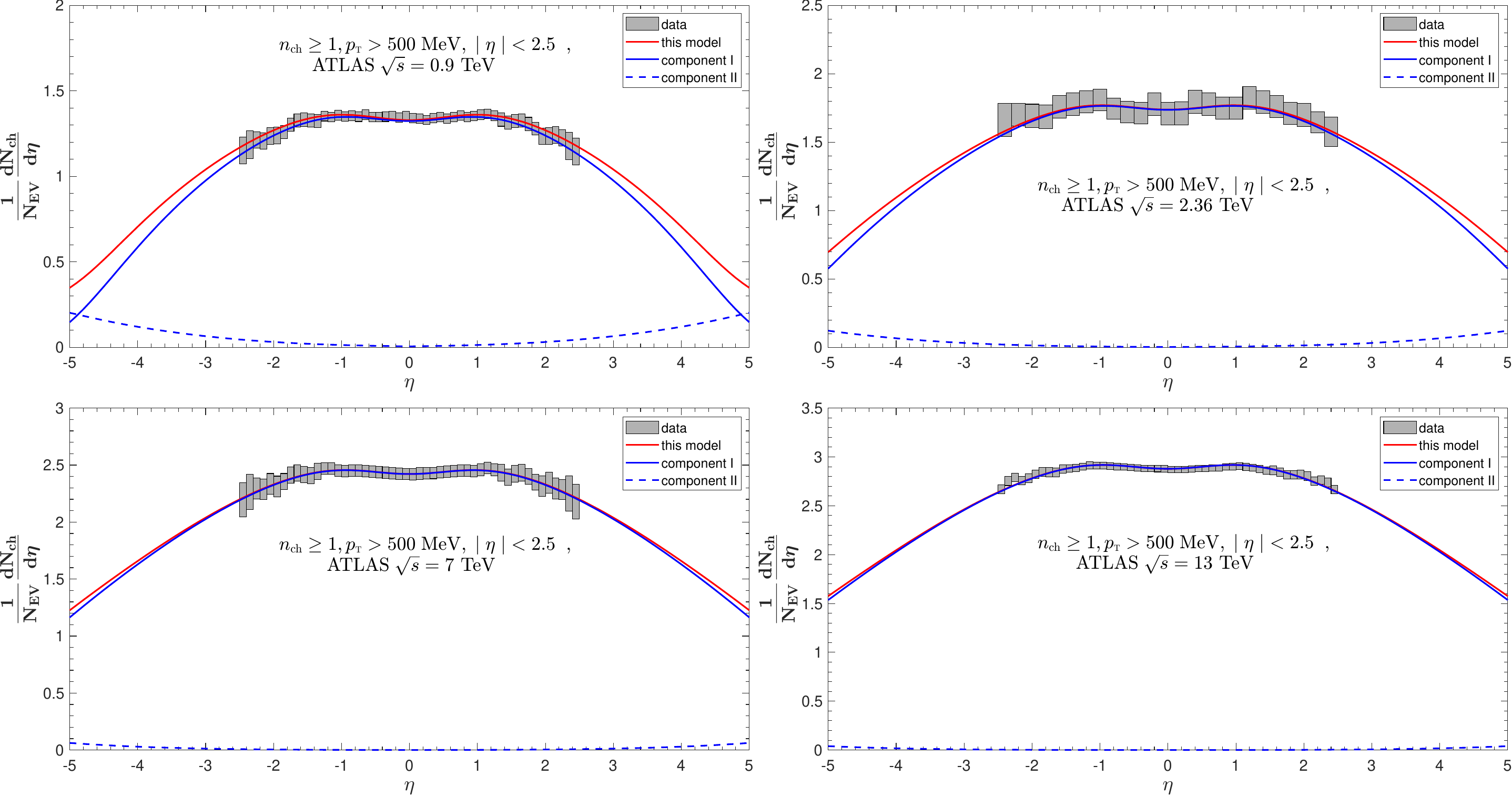}
    \caption{Comparison of the scaled two\text{-}component model predictions with ATLAS data($n_{ch}\ge1,p_{t}>500 \ \mathrm{MeV}$) for 
	$(1/N_{\mathrm{ev}}) \cdot (\mathrm{d}N_{\mathrm{ch}}/\mathrm{d}\eta) \quad \text{vs.} \quad \eta$ at 0.9, 2.36, 7, 13 {TeV} respectively. The model is normalized to match the data at
	$\eta=0$.}
   \label{fig:thismodel}
\end{figure*}

Figure~\ref{fig:thismodel} displays the model results. In the region $|\eta| < 2.5$, the model shows good agreement with the experimental data. As can be seen from Fig.~\ref{fig:thismodel}, the contribution from Component~II in the region $|\eta| < 2.5$ is small and becomes nearly negligible as the center-of-mass energy increases to $\sqrt{s} = 7~\mathrm{TeV}$. This behavior is consistent with the prediction reported in Ref.~\cite{ruan2010particle}, reflecting the dilution of valence-quark contributions with increasing collision energy in the central region.

\begin{figure*}[t]
    \centering
    \includegraphics[width=\textwidth]{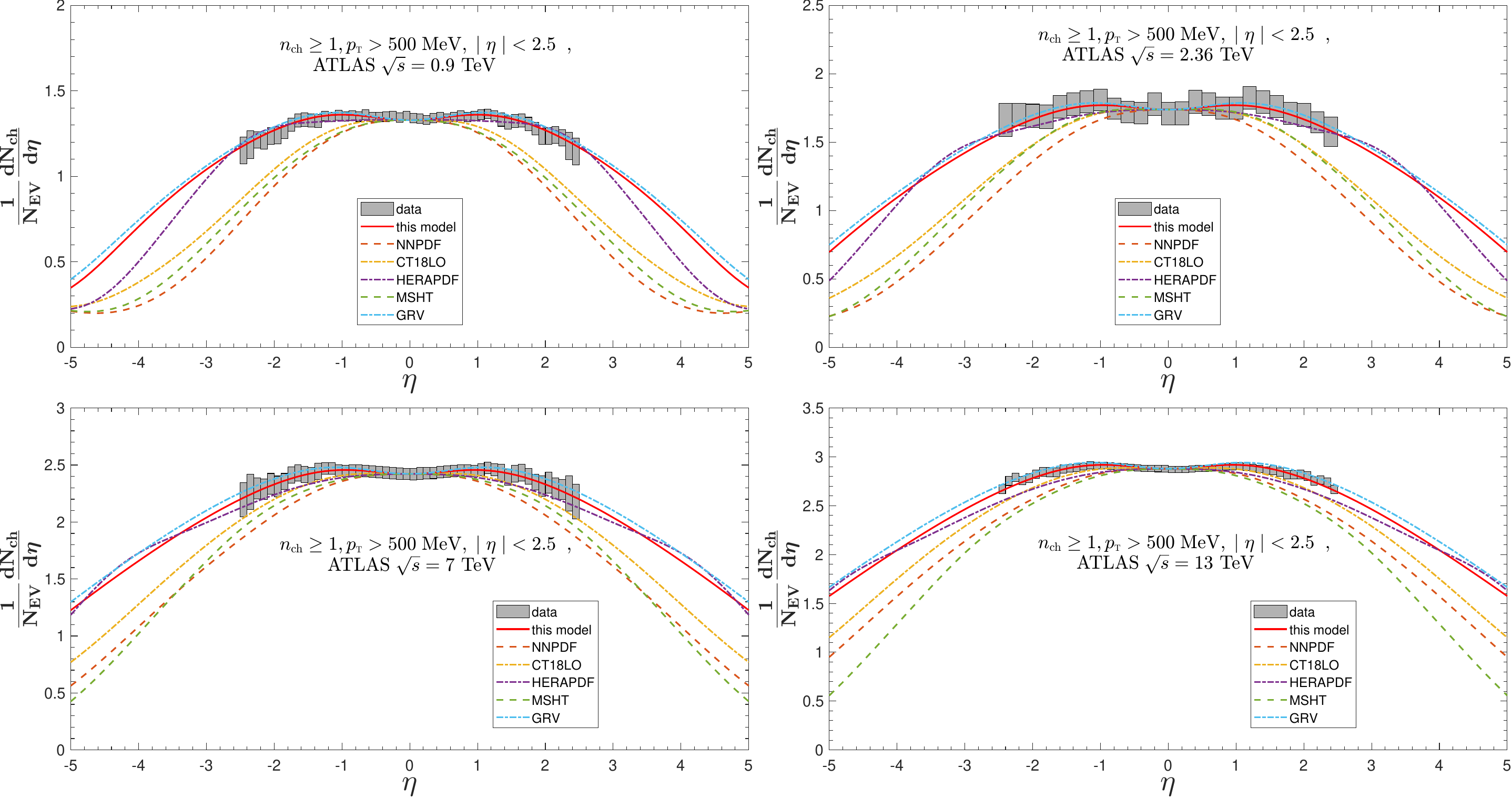}
    \caption{Predictions of Charged-Particle Pseudorapidity Spectra in pp Collisions at 0.9, 2.36, 7, 13 {TeV} respectively.
     Using Different PDF Sets, 
	compared to ATLAS Measurements. }
    \label{fig:VariedPDFs}
\end{figure*}

\begin{figure*}[t]
    \centering
    \includegraphics[width=\textwidth]{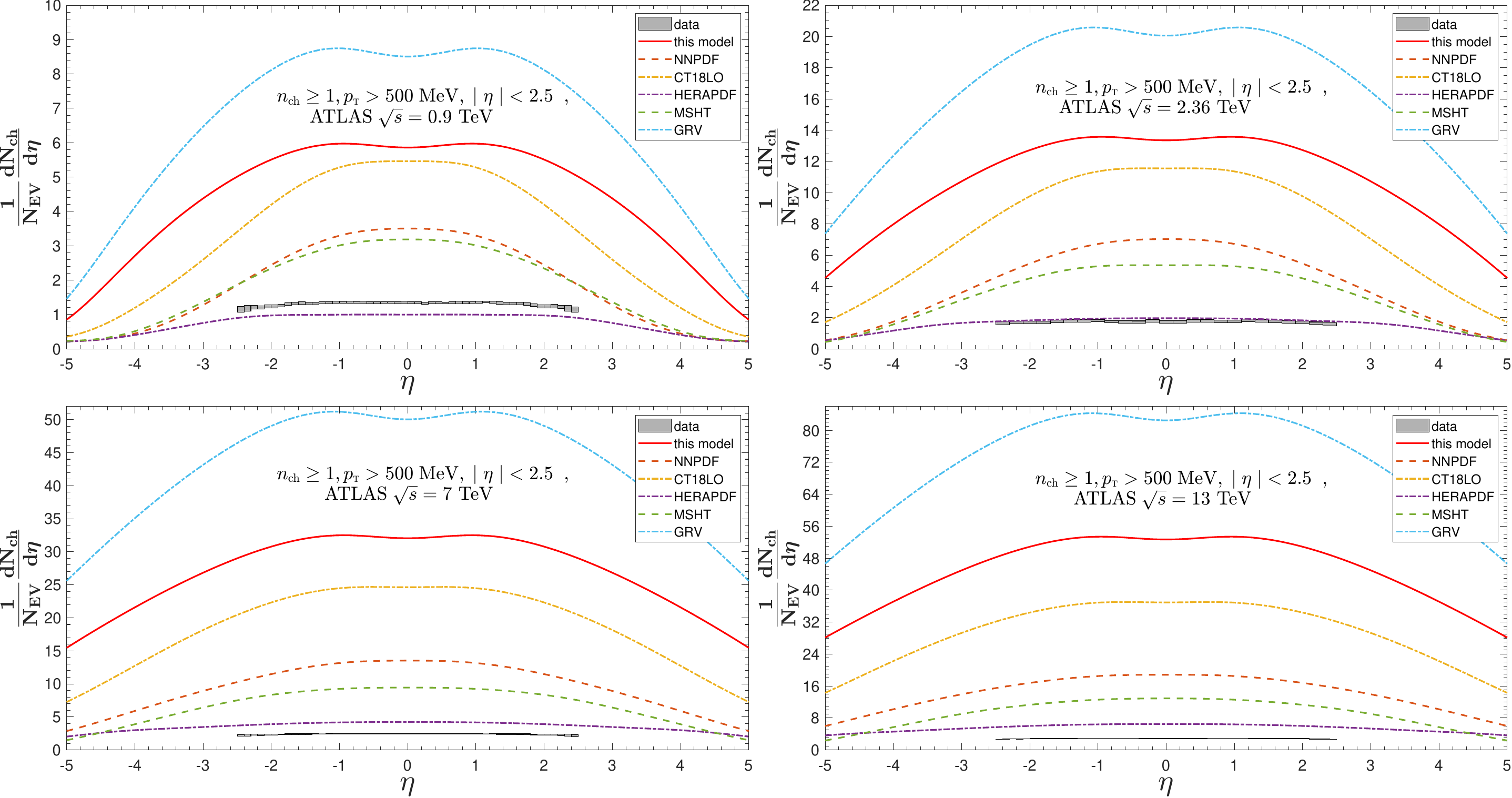}
    \caption{Predictions of Charged-Particle Pseudorapidity Spectra in $pp$ Collisions, as in Fig.~\ref{fig:VariedPDFs}, but calculated with the constant $C(\sqrt{s}) = 1$.}
    \label{fig:nocs}
\end{figure*}

\begin{figure*}[t]
    \centering
    \includegraphics[width=\textwidth]{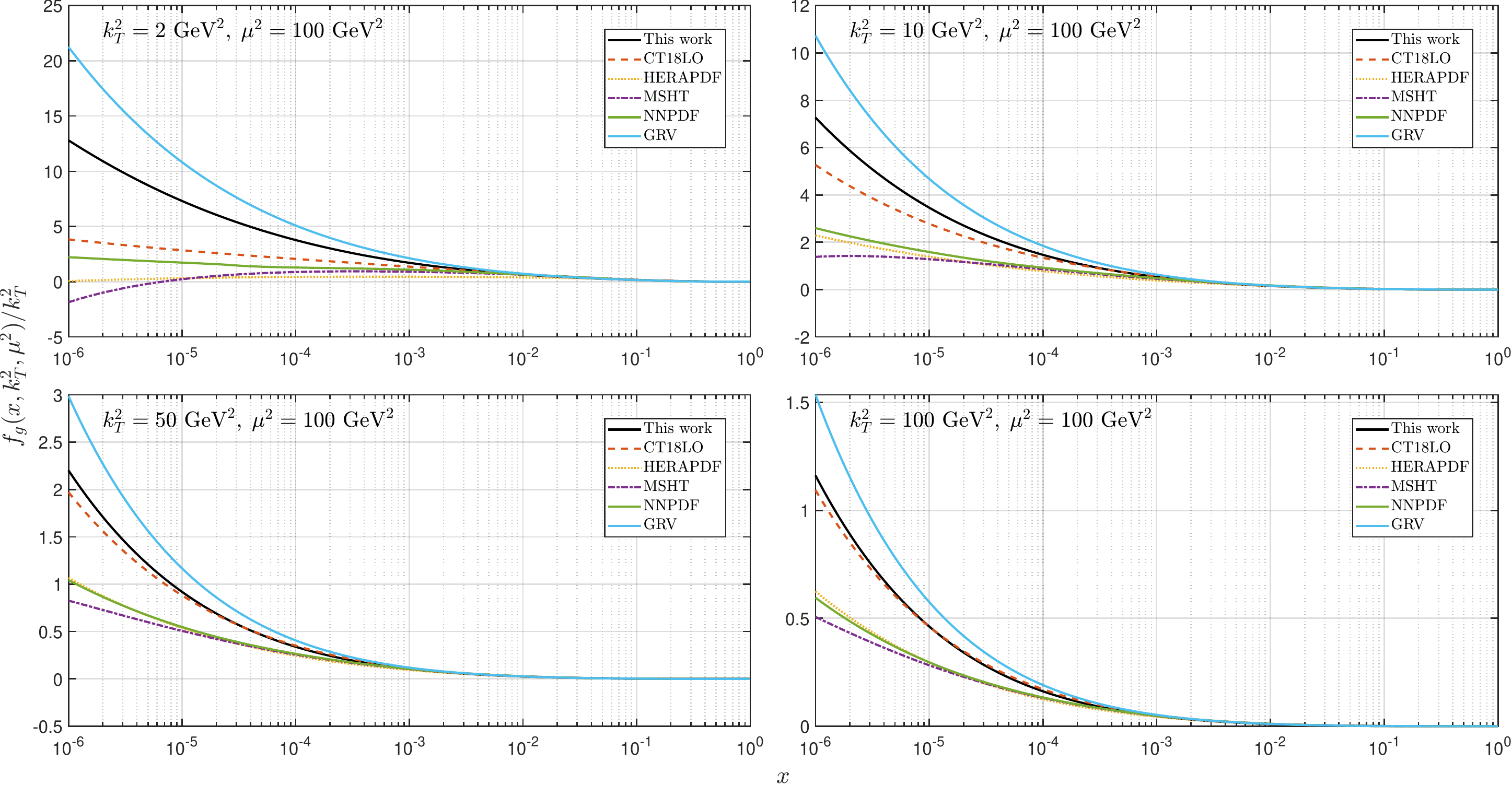}
    \caption{Unintegrated gluon distribution $f_g(x,k_T^2,\mu^2)/k_T^2$ at $\mu^2=100 \ \mathrm{GeV^2}$, comparing with 
    the result of CT18~\cite{yan2023ct18}, HERAPDF~\cite{collaborations2015combination}, MSHT~\cite{bailey2021parton}, NNPDF~\cite{ball2015parton} 
    and GRV~\cite{gluck1998dynamical}}
    \label{fig:fg_comparison}
\end{figure*}

\subsection{Comparing with other PDF sets}

In Fig.~\ref{fig:VariedPDFs}, we compare the results obtained using several different PDF inputs, including CTEQ~\cite{yan2023ct18}, MSHT~\cite{bailey2021parton}, NNPDF~\cite{ball2015parton}, HERAPDF~\cite{collaborations2015combination}, and GRV~\cite{gluck1998dynamical}. The corresponding unintegrated parton distribution functions are constructed using the same KMR prescription, without including higher-twist corrections. As shown in the figure, within the kinematic range covered by the experimental data, our results and those obtained with the GRV PDFs are generally in good agreement with the measurements. In contrast, the spectra predicted using the NNPDF, CTEQ, and MSHT sets are systematically narrower than the experimental distributions. The HERAPDF set provides a good description of the data at $\sqrt{s} = 0.9~\mathrm{TeV}$ but shows increasing deviations at higher collision energies. Among all the PDF sets considered, the GRV-based result yields the largest magnitude.
\clearpage
\onecolumngrid

\begin{center}
\captionof{table}{Longitudinal momentum fractions at different pseudorapidities. 
$x_1$,$x_2$ are defined in Eq.~\eqref{eq:gluonfusion}.}
\label{tab:x-fractions-eta}

\vspace{0.6ex}

\footnotesize
\setlength{\tabcolsep}{4.8pt}
\renewcommand{\arraystretch}{0.92}

\begin{tabular}{c *{5}{c}}
\toprule
$\sqrt{s}$ (TeV) 
& \multicolumn{5}{c}{Pseudorapidity $\eta$} \\
\cmidrule(lr){2-6}
& 0          & \multicolumn{2}{c}{2.5}          & \multicolumn{2}{c}{5.0} \\
\cmidrule(lr){2-2} \cmidrule(lr){3-4} \cmidrule(lr){5-6}
& $x$        & $x_1$           & $x_2$           & $x_1$           & $x_2$ \\
\midrule
0.9  & $1.11 \times 10^{-3}$ & $1.35 \times 10^{-2}$ & $9.12 \times 10^{-5}$ & $1.65 \times 10^{-1}$ & $7.49 \times 10^{-6}$ \\
2.36 & $4.24 \times 10^{-4}$ & $5.16 \times 10^{-3}$ & $3.48 \times 10^{-5}$ & $6.29 \times 10^{-2}$ & $2.86 \times 10^{-6}$ \\
7    & $1.43 \times 10^{-4}$ & $1.74 \times 10^{-3}$ & $1.17 \times 10^{-5}$ & $2.12 \times 10^{-2}$ & $9.63 \times 10^{-7}$ \\
13   & $7.69 \times 10^{-5}$ & $9.37 \times 10^{-4}$ & $6.31 \times 10^{-6}$ & $1.14 \times 10^{-2}$ & $5.18 \times 10^{-7}$ \\
\bottomrule
\end{tabular}

\end{center}

\twocolumngrid

It should be emphasized that the curves shown in Fig.~\ref{fig:VariedPDFs} do not directly reflect the relative normalizations of the different PDF sets, since they are multiplied by different coefficients $C(\sqrt{s})$ (see Eq.~\ref{eq:mult_I}) when compared with experimental data. To illustrate the intrinsic differences among the PDF sets, we set $C(\sqrt{s}) = 1$ for all cases, and the corresponding results are shown in Fig.~\ref{fig:nocs}. From this comparison, it is evident that the GRV result is the largest, whereas the HERAPDF result is the smallest and lies closest to the experimental data. These substantial differences among the predictions can be traced back to variations in the underlying gluon distributions, particularly in the small-$x$ region.

In Fig.~\ref{fig:fg_comparison}, we further compare the corresponding unintegrated gluon distribution functions obtained from the different 
PDF sets. Significant discrepancies among the distributions are observed in the region $x < 0.001$. In the present analysis, we consider 
proton--proton collisions at center-of-mass energies of 0.9, 2.36, 7, and 13~TeV. Within the pseudorapidity range $|\eta| < 2.5$ covered by 
the experimental data, the relevant parton momentum fractions predominantly fall in the interval $0.00001 < x < 0.001$, as summarized in 
Table~1. Here $x_1$ and $x_2$ are evaluated from Eq.~\eqref{eq:gluonfusion} with $p_t=1 \ \mathrm{GeV}$. 
Consequently, Fig.~\ref{fig:fg_comparison} directly illustrates the differences among the various PDF sets in the kinematic region 
most relevant to the present study.

Considering that the coefficient $C(\sqrt{s})$ cannot yet be determined from first principles, we disregard the relative normalization of the results and focus instead on the shape of the distributions. Within this perspective, our model demonstrates that the experimental data impose stringent constraints on the shape of the gluon distribution in the small-$x$ region. As shown in Fig.~4, the gluon distribution adopted in this work provides the best overall agreement with the data, followed by the GRV and HERAPDF sets.
According to equation (5), it can be determined that over a wide range of $x$, the relative magnitudes of these three sets of gluon distributions match the experimental data.

\section{Conclusion}

In this work, we have investigated the charged-particle pseudorapidity density $(1/N_{\mathrm{ev}}) \cdot (\mathrm{d}N_{\mathrm{ch}}/\mathrm{d}\eta)$ in high-energy $pp$ collisions within a two-component framework that incorporates both integrated and unintegrated parton distribution functions. The collinear PDFs are evolved using the modified DGLAP (MD-DGLAP) equations and subsequently transformed into unintegrated PDFs via the KMR prescription. A detailed comparison with experimental data from the ATLAS Collaboration shows that, despite its minimal and transparent structure, the model successfully reproduces the overall shape of the measured multiplicity distributions over a wide kinematic range. Comparisons with predictions based on other commonly used PDF sets are also presented.

The main conclusions of this study can be summarized as follows.

(1) Although the model formally consists of two components, particle production is dominated in practice by the first component in the energy region considering in this work. In particular, gluon--gluon fusion is identified as the leading mechanism for charged-particle production in the central rapidity region of $pp$ collisions at center-of-mass energies $\sqrt{s} \geq 900~\mathrm{GeV}$. This highlights the central role of gluon dynamics in the small-$x$ regime at high energies.

(2) The results provide strong support for the internal consistency and phenomenological viability of the MD-DGLAP evolution scheme. The specific initial PDFs employed in this work exhibit a broader dynamical range and greater flexibility than several widely used PDF sets. Moreover, the KMR prescription proves to be an effective method for constructing unintegrated PDFs, enabling a unified description of particle production in both the central and fragmentation regions within a single theoretical framework.

(3) Significant differences are observed among the gluon distributions of commonly used PDF sets in the small-$x$ region. Proton--proton collision data, when analyzed within the present model, offer a sensitive means of discriminating between these PDFs and assessing their phenomenological validity.

In summary, this work presents a simple yet effective approach for describing charged-particle multiplicities in high-energy $pp$ collisions. It provides a useful baseline for future studies of other hadronic observables and offers a direct framework for testing and constraining parton distribution functions using the high-energy collider data.

\begin{acknowledgments}
This work is supported by the National Natural Science Foundation of China (grant No. 11851303)
\end{acknowledgments}

\appendix
\section{Splitting Functions and Input Distributions}
\label{app:splitting}

The unregularized DGLAP splitting kernels appearing in Eqs.~(\ref{eq:1}) and~(\ref{eq:2}) are given by
\begin{align}
P_{gg}(z) &= 2 C_2(G) \left[ z(1-z) + \frac{1-z}{z} + \frac{z}{1-z} \right], \\
P_{gq}(z) &= C_2(R) \frac{1 + (1-z)^2}{z}, \\
P_{qq}(z) &= C_2(R) \frac{1 + z^2}{1-z}, \\
P_{qg}(z) &= \frac{1}{2} \left[ z^2 + (1-z)^2 \right],
\end{align}
with color factors $C_2(G) = N = 3$ and $C_2(R) = (N^2 - 1)/(2N) = 4/3$.

The recombination functions are
\begin{align}
P_{gg \to g}(z) 
&= \frac{9}{64} \frac{1}{x} (2 - z) \Bigl[ 
   72 - 48z + 140z^2 
\\&\qquad
   - 116z^3 + 29z^4 
   \Bigr],
\\[1.2ex]
P_{gg \to q}(z) 
&= \frac{1}{48} \frac{1}{x} z (2 - z)^2 (18 - 21z + 14z^2).
\end{align}

The input distributions for the MD-DGLAP evolution, based on GRV98LO but with modifications, are as follows. The initial valence 
quark and gluon densities are
\begin{align}
x u_v(x, \mu_0^2) &= 1.239 x^{0.48} (1-x)^{2.72} (1 - 1.8 \sqrt{x} + 9.5 x), \\
x d_v(x, \mu_0^2) &= 0.614 (1-x)^{0.9} x u_v(x, \mu_0^2), \\
x g(x, \mu_0^2) &= 17.47 x^{1.6} (1-x)^{3.8}.
\end{align}
The input sea quark distribution is modified to
\begin{align}
x\, s(x,\mu_0^2)
&= 2x\,[\bar{u} + \bar{d}] \nonumber \\
&= 0.9\, x^{0.01}
   \bigl(1 - 3.6\sqrt{x} + 7.8x\bigr)
   (1-x)^{8.0}
\end{align}

with the starting scale adjusted from $\mu_0^2 = 0.26$ GeV$^2$ to $\mu_0^2 = 0.34$ GeV$^2$. 
For further details, see Ref.~\cite{ruan2009prediction}.

\nocite{*}

\bibliography{apssamp}

\end{document}